# Crystallographic image processing for scanning probe microscopy

**P. Moeck**

Department of Physics, Portland State University, Portland, OR 97207-0751, U.S.A.

Scanning probe microscopy (SPM) images of regularly arranged spatially periodic objects can be processed crystallographically. The resulting information may be used to remove from the SPM image distortions that are due to a "less than perfect" imaging process. The combined effects of these distortions result in a point spread function that gives a quantitative measure of the microscope's performance for a certain set of experimental conditions. On the basis of highly symmetric "calibration samples", the point spread function of the microscope may be extracted and utilized for the correction of SPM images of unknowns that were recorded under essentially the same experimental conditions. We concentrate in this paper on more theoretical aspects of our method. A "blunt" scanning tunnelling microscopy (STM) tip that consists of multiple "mini-tips" with electron orbital dimensions may be "symmetrized" on the basis of prior knowledge on the plane symmetry of a two-dimensional periodic array. This is illustrated with the crystallographic processing of a STM image of a regular array of fluorinated cobalt phthalocyanine molecules on graphite and backed up conceptually by simple simulations.

**Keywords** scanning probe microscopy; scanning tunnelling microscopy; 2D periodic symmetry; point spread function

## 1. Introduction

The defining features of traditional[1] scanning probe microscopes (SPMs) are a very fine "probe" that is scanned in two dimensions (2D), in very fine steps, very close to the surface of a sample, and a "probe-sample interactions signal" that is recorded at each scanning increment. This signal can be digitized and displayed as a function of the magnified scanning steps. The result is a 2D-image of the probe-sample interactions. Just like any other image, a SPM image can be subjected to image processing routines in order to quantify the information that it contains.

The SPM imaging process results unavoidably in a convolution of a scanning probe tip function that includes its geometry (and possible asymmetry) with the respective function of the features of the sample surface that includes their geometry (and possible 2D-periodic symmetry). It is, therefore, important that the physical dimensions of this tip are much smaller than the surface features it is supposed to probe.

Conversely, if the scanning probe tip is laterally wider than the surface features, the SPM image will be composed of representations of the tip as probed by the sample surface features [1]. Deconvolution of the scanning probe tip size and geometry, i.e. so called "tip dilation", in order to recover more genuine features of the sample surface is common practice in SPM imaging [2]. Such deconvolutions are particularly important for atomic force microscopy (AFM) imaging with inexpensive commercial equipment at resolutions in the tens of nanometre range. Scanning tunnelling microscopy (STM) imaging relies ideally on a single protruding orbital of a single atom at an atomically sharp apex of the scanning probe tip [3], but a blunt tip may also be imaged inadvertently by sample surface atoms. Such a blunt STM tip may be composed of many mini-tips, i.e. atomic protrusions that act as tunnelling sites simultaneously, and may be much wider laterally than the spatial extension and separation of the individual atomic surface features it is supposed to image.

Although the scanning probe microscopy community has been analyzing regular 2D-periodic arrays right from the invention of the various types of SPMs, it did not occur to many members of this community (who lacked thorough crystallographic training) to utilize crystallographic image processing (henceforth abbreviated as CIP) in order to help quantifying the information that is contained in the respective 2D images. This is completely understandable as most members of this community concentrated on improving the various imaging techniques and pushed them to higher and higher resolutions while increasing the meaningful magnifications. As a result, smaller and smaller entities, e.g. atoms within individual organic molecules could be imaged [4] (even with AFM when atomically sharp tips were utilized in expensive user-build research equipment), but with signal to noise ratios that may still be improved.

Crystallographers have traditionally taken a different approach to atomic resolution. The kinematic scattering of photons, electrons and neutrons can be modelled by a Fourier transform. While the Fourier transform of a single molecule is a continuous function in Fourier space, the Fourier transform of a regular array of molecules is a discrete series of Fourier coefficients. The dimensionality and real space lattice of an array of molecules determines the dimensionality and reciprocal space lattice of its Fourier transform. Strictly speaking, the array needs to range to infinity in order to employ Fourier techniques in the mathematically exact manner. Some finite section of an infinitely large array is, however, typically sufficient in order to begin benefiting from the crystallographic approach that utilizes finite and discrete Fourier series.

---

[1] "Traditional" refers here to SPMs that have just one "kind" of a probe-sample interactions signal attached to each 2D scanning increment. Non-traditional SPMs are defined by having two or more such kinds of signals attached to each 2D scanning increment. Examples of non-traditional SPMs are spin-polarized STMs and critical dimension SPMs. Utilizing "black-white" and "colour symmetries" (e.g. Shubnikov AV, Below NV. et al., *Colored Symmetry*, Ed. W. T. Holser, Pergamon Press, Oxford, 1964; Vainshtein BV. *Fundamentals of Crystals, Symmetry and Methods of Structural Crystallography*, 2$^{nd}$ edition, Springer-Verlag, Berlin, Heidelberg, 1994) as group theoretical basis, all concepts we mention in this book chapter can be straightforwardly expanded to the crystallographic processing of spatially periodic images that were recorded with such microscopes.

CIP originated some 40 years ago with the electron crystallography community and is partially based on earlier ideas from the X-ray crystallography and optical transform communities. Its underlying theory is independent of the source of the patterns. All of its concepts can, therefore, be straightforwardly extended to scanning probe microscopy. One may define CIP simply as being concerned with categorizing, specifying, and quantifying 2D (or 1D) spatially periodic, perfectly long-range ordered patterns in direct space. Similar to the theories of 3D X-ray or electron diffraction crystallography, the theory behind CIP applies also to finite spatially and temporally averaged patterns.

The averaging over many periodic motifs of which a few may be "defective" results in the determination of a structure that is considered as being essentially free of defects but close to the ideal (defect free) structure [5]. The defects themselves may be either local, e.g. vacancies at the repeating motif level, or concern the whole pattern, e.g. overall 2D geometric distortions and the effects of a possible scanning probe tip bluntness. The real pattern is then a combination of the ideal pattern with some additional quantitative measure of the effects of the defects. This is all analogous to the concepts of quantified ideal and real crystal structures in materials science [5]. A beneficial side effect of the averaging over many periodic motifs is an enhancement of the signal to noise ratio of the imaging process.

In this book chapter, we will ignore local defects and noise entirely and consider defects that affect the whole pattern as being caused by the imperfections of the SPM imaging process alone. These imperfections result then in a prevalent point spread function of the microscope for a set of experimental imaging conditions including the effective scanning probe tip, which may be blunt. Obtaining an estimate of this function and correcting for the effects of this function (including the symmetrization of a blunt scanning probe tip) become then parts of CIP.

In order to enable the utilization of CIP by other members of the scanning probe microscopy community, we concentrate in the main part of this paper on more theoretical aspects of our method. This book chapter may, therefore, be considered as constituting a brief tutorial on the theory and praxis of CIP as applied to SPMs.

Utilizing CIP, we will demonstrate in the second part of this paper briefly the symmetrization of a blunt STM tip on the basis of (presumed) prior knowledge about fluorinated cobalt phthalocyanine molecules that are arranged regularly on graphite in a 2D spatially periodic manner. Such a scanning probe tip symmetrization is finally backed up conceptually by two simple simulations.

## 2. Theory and praxis of crystallographic image processing

### 2.1 Background and notation

Within the weak phase object approximation of transmission electron microscopy, CIP is widely employed in electron crystallography to aid the extraction of structure factor amplitudes and phase angles from high resolution phase-contrast transmission electron microscopy images of crystalline materials [6, 7]. It is also used for the correction of these images for the effects of the phase-contrast transfer function, two-fold astigmatism, beam tilt away from the optical axis of the microscope, and sample tilt away from low indexed zone axes. The product of all of the "respective aberration functions" may be considered as the Fourier transform of the point spread function of the microscope under the prevalent experimental conditions. The effect of this point spread function can be removed from the image by the enforcing of a correctly determined 2D symmetry on the high resolution phase-contrast transmission electron microscopy image.

The underlying symmetry quantification and enforcing principles of CIP are completely general. The serial nature of the acquisition of a SPM image is thereby utterly unimportant. This has been demonstrated by the recent application of popular CIP software from the electron crystallography community [8] to images that were recorded either in the Z-contrast mode of scanning transmission electron microscopy [9] or in a STM [10, 11]. The generality of the CIP procedures in 2D is due to the fact that there are just 17 mathematically defined space symmetry groups for 2D-periodic entities that may exist in nature [6, 7, 12-14]. (These groups are also known as "one-sided plane groups" or simply "wallpaper" groups.) Sixteen of these plane groups possess more point symmetry elements than the identity element. We will refer to these groups below as higher symmetric plane groups. (The crystallographic processing of images of 1D spatially periodic objects that extend over two dimensions is, on the other hand, based on the 7 frieze groups [15] that are also known as "one-sided band" (or border) groups, but will not be elaborated further here.)

In this paper, we will use only the international (Hermann-Mauguin) notation for both point and plane symmetry groups [12]. Due to their geometrical importance, at least seven other notations have been proposed for the plane groups in the past [13]. A slightly modified notation that also considers non-standard orientations of mirror or glide lines for plane groups *pm*, *pg*, *cm*, and *p2mg* (that all posses rectangular lattices) is popular in the electron crystallography community that solves and refines inorganics [6]. In that notation, the mirror or glide lines can be oriented either perpendicular to the x-axis (i.e. in their standard orientation [12]) or perpendicular to the y-axis.

The Schoenflies notation of 2D point groups is popular with chemists that have interests in the spectroscopy of molecules. For a conversion from the latter notation to the Hermann-Mauguin 2D point group notation, see ref. [5]. (There is no Schoenflies notation for the plane symmetry groups.) The electron crystallography community that deals with pseudo-2D membrane protein crystals [7] prefers sometimes the international notation of the 80 layer groups[2] [15].

### 2.2 Quantifying deviations from plane symmetries and deciding on the most likely plane group

In order to enforce the most likely plane symmetry on an experimental image and remove geometric distortions, defects and noise in the process, one must first be able to quantify the deviations of the image from the 16 higher symmetric plane symmetries and then make the correct decision on which plane group to enforce. Popular CIP software [8] utilizes ideas from diffraction-based electron crystallography[3] for the quantification of the deviations of an experimental image from the plane symmetries of the higher symmetric groups. The decision on which plane group is the most likely is currently somewhat subjective because supergroup/subgroup relations [19] are not considered in a quantitative manner.

Provided that an image is periodic in 2D, a discrete set of Fourier coefficients of the image intensity represents the image in reciprocal space. The amplitude *(A)* and phase angle *(φ)* of the Fourier coefficients of the image intensity have to obey certain symmetry relations and restrictions in order to belong to one of the 16 higher symmetric plane groups [6, 7]. Centred lattices and glide lines lead to so called "systematic absences" (also referred to as "extinctions" in reciprocal space, i.e. certain Fourier coefficients that must be zero if the 2D image is to belong to the respective plane symmetries [6, 7, 12].

In order to quantify deviations from the 16 higher symmetric plane groups, the electron crystallography community uses Fourier coefficient amplitude *($A_{res}$)* and phase angle *($\varphi_{res}$)* residuals as defined by the relations:

$$A_{res} = \frac{\sum_{H,K} \left||A_{obs}(H,K)| - |A_{sym}(H,K)|\right|}{\sum_{H,K} |A_{obs}(H,K)|} \quad \text{in \%} \quad (1a) \quad \text{and} \quad \varphi_{res} = \frac{\sum_{H,K} w(H,K) \cdot |\varphi_{obs}(H,K) - \varphi_{sym}(H,K)|}{\sum_{H,K} w(H,K)} \quad \text{in degrees} \quad (1b),$$

where the subscripts stand for *obs*erved and *sym*metrized, *w* is a relative weight (that is in popular software [8] set proportional to $A_{obs}$), and the sums are taken over all Fourier coefficient labels *H* and *K* [6, 7]. Note that we are in the process of redefining[3] these residuals for the utilization of CIP in scanning probe microscopy in order to make the subsequent decision to which plane group an experimental image most likely belongs less subjective. In the future we will develop more quantitative methods and criteria[4] to substantiate these decisions.

Since certain plane groups require certain 2D lattices [12], it makes sense to determine the overall type of the lattice (oblique, rectangular, square, or hexagonal) first and then to calculate the amplitude and phase angle residuals only for the respective plane groups and all of their subgroups in lower symmetric lattices. The determination of the lattice type is based on the reciprocal lattice as obtained from the Fourier transform of the image intensity. Rectangular lattices can be either primitive (i.e. contain one motif) or centred (i.e. contain two motifs).

When an observed Fourier coefficient amplitude is not symmetry related to other observed Fourier coefficient amplitudes (except to that of its Friedel pair[5]), the symmetrized amplitude is set equal to its observed counterpart. Otherwise the symmetrized Fourier coefficient amplitude of *(H,K)* is averaged over all *(n)* observed symmetry-related Fourier coefficient amplitudes with indices *(H',K')* (including the observed amplitude of the original Fourier coefficient):

$$A_{sym}(H,K) = \frac{\sum_n |A_{obs}(H',K')|}{n} \quad (2a), \text{ see ref. [6]}.$$

---

[2] Regular arrays of chiral proteins that form membranes which are only one unit cell thick (in the direction of the transmitting electron beam) are often considered as "pseudo-2D crystals" and referred to by the layer group variant of their space group symbols by the electron crystallography community that deals with membrane protein crystals [7]. There are also 17 groups of these "pseudo-2D membrane protein crystal symmetries" that belong to the "two-sided plane groups" and possess symmetry related ½+z and ½−z positions. These positions are due to two-fold rotation (*2*) and screw axes (*$2_1$*) which are perpendicular to the membrane thickness and located half way between the front and back sides of the membranes. Often the international layer group notation [15] of these "pseudo-2D groups" is used even when the referred symmetry is strictly in 2D, i.e. when one considers 2D projections where the only possibly z-coordinate is zero and one should per definition use the notation of the plane groups [12]. The conversions between both notations are straightforward, i.e. 2 ↔ *m*, *$2_1$* ↔ *g*, where *m* stands for a mirror and *g* for a glide line.

[3] Relation (1a) is currently defined analogously to the so called "internal symmetry factor" of diffraction based electron crystallography [16]. Provided that the correct 2D or 3D space group has been employed, this factor allows for an assessment of the "quality" of a 2D or 3D dataset. Similar residuals are used in X-ray crystallography. There have been attempts to redefine such residuals in order to take the multiplicity of the X-ray reflections into account [17, 18]. We plan to follow these lines of thought and to redefine relation (1a) so that the multiplicity of each set of symmetry equivalent Fourier coefficient amplitudes is properly weighted. We also plan to redefine relation (1b) in a manner so that different sets of symmetry equivalent Fourier coefficient phase angles are weighted differently.

[4] It should be possible to define and utilize geometric Akaike criteria [20] in order to decide which plane group is the most likely within sets of families of plane groups that are in supergroup/subgroup relationships [21]. Cross-correlations in direct space may also be utilized for this purpose.

[5] In X-ray crystallography, Friedel pairs are formed by the reflections (hkl) and (-h-k-l). Under ideal kinematic diffraction conditions, which can be modelled by a Fourier transform, the intensity of both reflections are exactly the same. This is generally known as Friedel's law, from which one can derive for 2D Fourier coefficient amplitudes $|A(H,K)| = |A(-H,-K)|$ and $\varphi(H,K) = -\varphi(-H,-K)$ for their phase angles.

When an observed Fourier coefficient phase angle is not symmetry related to other observed Fourier coefficient phase angles (except by Friedel's law[5]), the symmetrized phase angle is set equal to its observed counterpart. Otherwise the symmetrized phase angle is averaged over all observed symmetry-related Fourier coefficient phase angles (including the observed phase angle of the original Fourier coefficient) according to the relation:

$$\varphi_{sym}(H,K) = \arctan\left[\frac{\sum_j w^j s^j \sin\{\varphi_{obs}^j(H',K')\}}{\sum_j w^j s^j \cos\{\varphi_{obs}^j(H',K')\}}\right] + \begin{cases} 0° & [if \sum_j w^j s^j \cos\{\varphi_{obs}^j(H',K')\} > 0] \\ 180° & [if \sum_j w^j s^j \cos\{\varphi_{obs}^j(H',K')\} < 0] \end{cases} \text{ in degrees} \quad (2b),$$

where the sums are over all symmetry related phase angles (including the observed Fourier coefficient phase angle of the original Fourier coefficient), $w^j$ is a weighting factor, and $s^j = 1$ if the observed and symmetrized phase angles are approximately the same or $s^j = -1$ if they differ by approximately 180°, see ref. [6]. This symmetry averaging procedure of the Fourier coefficient amplitudes and phase angles corresponds to "vector averaging" in the Gaussian plane of complex numbers.

While all phase angles have to be either 0° or 180° for the 10 "pseudo-centrosymmetric" plane symmetry groups (that possess two-fold rotation points as projections of 3D inversion centres), this restriction applies only to subsets of the Fourier coefficients for plane groups *pm*, *pg*, *cm*, *p3m1*, and *p31m*. Where required by the plane symmetry group, the respective symmetrized phase angle is finally set to 0° if this angle is within the range smaller than -90° or 90°. If this angle is larger than (or equal to) -90° or 90°, it is finally set to 180°, see ref. [6].

While the amplitudes of the Fourier coefficients are theoretically independent of the point about which a Fourier transform may be calculated, the phase angles of the Fourier coefficients are always defined with respect to the chosen origin in reciprocal space. One needs, therefore, to shift this origin to the crystallographic origin of the 16 higher symmetric plane groups [12]. Popular CIP software [8] does this by calculating the phase angle (and amplitude) residual(s) for each pixel in the 2D unit cell as determined in the previous 2D lattice determination step and by picking the lowest of these phase angle residuals (together with the respective amplitude residual) as representative of the respective group at the crystallographically defined origin. This so called "origin refinement procedure" identifies the one pixel (and all of its translation equivalent pixels) in the unit cell of the direct space 2D image about which the respective plane symmetry is the least broken. (For plane groups that do not contain a two-fold or higher-fold rotation point, these pixels will lie on straight lines.)

The amplitude residual is "trivial" for plane group *p2*, i.e. equal to zero and meaningless. This is because the only symmetry relation between the amplitudes of the Fourier coefficients of the image intensity in this group is $A(H,K) = A(-H,-K)$, which is a property of the Fourier transform[5] itself. While $\varphi(H,K) = -\varphi(-H,-K)$ is also a property of the Fourier transform[5], the phase angle residual for plane group *p2* is still meaningful because all phase angles are restricted by the pseudo-centrosymmetry of this plane group to be either 0° or 180°. For plane group *p1*, both residuals are trivial since there is only translation symmetry and the identity point symmetry element. Consequently, there are neither symmetry relations nor restrictions between the Fourier coefficients of the image intensity.

Provided that an image possesses a sufficient number of spatial periodicities in direct space, i.e. possess a sufficient number of Fourier coefficients in reciprocal space, the amplitude and phase angle residuals of the Fourier coefficients of the image intensity are likely to be different for each of the 14 higher symmetric plane groups that are based on primitive lattices. Note that while relations (1a) and (1b) are valid for all 16 higher symmetric plane groups and the vector averaging in the Gaussian plane of complex numbers follows the same rules, i.e. relations (2a) and (2b), different sets of Fourier coefficients tend to be averaged and phase angle restricted (when restrictions apply) for the 14 higher symmetric plane groups that are based on primitive lattices. The Fourier coefficient amplitude and phase angle residuals provide, thus, powerful quantitative measures to identify the most likely plane symmetry group that a 2D-periodic image may possess when geometric distortions, defects and noise are removed.

In addition, the popular CIP software CRISP (but not the software 2dx) [8] defines a so called *Ao/Ae* ratio for those 6 plane groups that possess systematic absences [6, 12]. This ratio is defined as the amplitude sum of the Fourier coefficients that are forbidden by the symmetry but were nevertheless observed (*Ao*), i.e. so called "forbidden Fourier coefficients", divided by the amplitude sum of the observed Fourier coefficients (*Ae*) that are allowed by the plane symmetry. An *Ao/Ae* ratio larger than zero, thus, means that plane group forbidden Fourier transform coefficients are actually present with an amplitude greater than zero. This makes this ratio (for the six plane groups to which it is applicable) an additional measure to quantify deviations of the symmetry of an image and helps in the identification of the most likely plane symmetry group that a 2D-periodic image may possess when geometric distortions, defects and noise are removed.

Note that systematic absences are due to both glide lines and cell centring [6, 12]. The *Ao/Ae* ratio, thus, allows one to distinguish the plane groups *cm* and *c2mm* from the plane groups *pm* and *p2mm*, respectively. One can, thus, distinguish between all 16 higher symmetric plane groups on the basis of a combination of the *Ao/Ae* ratio with the Fourier coefficient amplitude and phase angle residuals.

For an exact adherence of an image to one of the 16 higher symmetric plane groups, both the amplitude and phase angle residuals need to be zero. In addition, for an exact adherence of an image to one of the 6 plane groups that possess systematic absences, the *Ao/Ae* ratio needs to be zero. Such an exact adherence of an experimental image to an abstract mathematical definition of symmetry is not to be expected. Roughly speaking, this would correspond to a measurement without any systematic and random errors.

Low Fourier coefficient amplitude and phase angle residuals, together with very low or zero *Ao/Ae* ratios (when the latter measure is meaningful) form the basis for the decision as to which plane group an experimental image most likely belongs. One also needs to take the multiplicity (*M*) of the general position (and of positions with higher site symmetries) into account. This multiplicity ranges from 2 to 12 and is the higher the more symmetry elements exist in the plane group. Supergroup/subgroup relations [12] must also be considered.

The general rule is that the most likely plane group is the one with the highest multiplicity of the general position within a set of supergroup/subgroup related plane groups that at the same time has reasonably low amplitude and phase angle residuals combined with very low or zero *Ao/Ae* ratios (when this ratio is meaningful). Typically, the phase angle residual is more valuable than the amplitude residual for considering which plane group an experimental image most likely belongs to[6]. These criteria are obviously somewhat subjective at present, but nevertheless used currently throughout the electron crystallography community. (As already mentioned above, we plan to develop more objective criteria to aide decisions to which plane group an experimental image most likely belongs and will also implement them into future CIP software [8].)

Frequently, the plane group to which an experimental image belongs is known in advance, see e.g. ref. [9], and the CIP procedure is mainly used to enhance the signal to noise ratio by plane symmetry averaging in reciprocal space and Fourier back transforming into direct space. Other cases of prior knowledge of the plane group of the sample surface features concern dedicated SPM calibration samples.

To summarize the main ideas of this section: the Fourier coefficient amplitude and phase angle residuals combined with the *Ao/Ae* ratios provide the means to quantify deviations of symmetry in 2D and allow for the currently somewhat subjective identification of the most likely plane group to which an experimental SPM image would belong when geometric distortions, defects and noise as well as a possible scanning probe tip bluntness were removed. It is important to stress that for the procedures to work quite unambiguously, the quantified deviations from the symmetries need to be small, i.e. that the distortions to the ideal SPM image must have been small in the first place.

### 2.3 Enforcing plane symmetries and reducing noise in the process

Plane symmetry enforcing on an SPM image consists of the Fourier back transforming of the symmetrized Fourier coefficients rather than of the observed Fourier coefficients of the image intensity. The symmetrized Fourier coefficients are obtained by relations (2a) and (2b), including the setting of the phase angles to 0° or 180° and the removal of forbidden Fourier coefficients (where applicable) for the crystallographic origin of the respective plane group as defined in ref. [12] and as obtained form the origin refinement procedure of CIP.

Because there are no point symmetry elements in plane group *p1* besides the identity element (*1*) and no restrictions on the geometry of the 2D lattice, there is no "genuine" plane symmetry enforcing procedure for this group. Fourier filtering (that is also known as optical filtering), e.g. ref. [22], may be considered as being conceptually equivalent to translation averaging over the periodic motif without enforcing any additional site symmetry. Since only the averaged translation symmetry will be enforced in a Fourier back-transform of non-symmetrized Fourier coefficients, one may consider Fourier filtering as being conceptionally equivalent to "quasi" *p1* plane symmetry enforcing. It is well known that a beneficial noise suppression results from the translation averaging over the periodic motif in an image. This noise suppression stems from the fact that noise is non-periodic.

In direct space, plane symmetry enforcing results in the averaging of the image intensity over all point symmetries as well as all translation symmetries. The averaging is done over the intensity of *all* symmetry related pixels of the *whole* image so that the respective point and translation symmetry conditions (and restrictions) are *all* fulfilled. An alternative view is that enforcing a higher symmetric plane symmetry (other than *pm*, *pg*, and *cm*) on an image results in the periodic motif being placed with its central rotation point (i.e. the projection of a rotation axis which is oriented perpendicular to the x-y plane) on the crystallographic origin and at all translation invariant positions of that plane group. In all cases, the respective point symmetry of the motif along with the respective site symmetries of all other points in the image and the translation symmetry of the motif are *all* enforced by the plane symmetry group enforcement procedure.

Combining point symmetry averaging with translation averaging to plane symmetry enforcing for the 16 higher symmetric groups results, thus, in "more averaging" for the same number (*N*) of periodic motifs. For all positions in the unit cell with site symmetry *1*, the *N* fold averaging is extended by a factor *M*, which represents the multiplicity of the general position of the respective plane group. One may also say that plane symmetry enforcing for the 16 higher symmetric groups means that one averages over all *N • M* asymmetric units in the *whole* image rather than over all *N* unit cells. The benefit of this "extended averaging" is obviously a better noise and (non-periodic) defect suppression than what can be achieved by translation averaging (also known as Fourier or optical filtering) over the same (*N*) number of unit cells alone.

---

[6] Sven Hovmöller of the University of Stockholm created a rule of thumb that states: *"Phase angles are at least twice as valuable for 2D image-based electron crystallography as amplitudes."* Henk Schenk of the University of Amsterdam commented along similar lines: *"Even when the structure factors are all set to one or taken from another crystal, the desired structure will show up in a Fourier summation, provided the correct phases have been used. However, wrong phases and correct amplitudes reveal no structure."* In: H. Schenk (Ed.) *Direct Methods of Solving Crystal Structures*, New York, NY: Plenum; 1991:1-8.

This leads to the alternative view that the enforcing of the correct plane symmetry results in the quite effective removal of defects, noise and geometric distortions including effects of multiple mini-tips from an experimental SPM image. The "effective sum" of all geometric distortions is considered to be caused by the prevalent point spread function of the microscope.

On the basis of the reciprocity principle of SPM imaging [3], plane symmetry enforcing on a SPM image can also be considered as being equivalent to scanning probe tip symmetrization. Utilizing Pierre Curie's symmetry principle [5, 23], (i.e. the symmetry of a crystal under the influence of a physical probe is the intersection of the symmetry group of the probe and the symmetry group of the crystal without the influence,) we can make the general statement that the point symmetry of the symmetrized scanning probe tip will be simply the highest site symmetry of the plane group that has been CIP enforced on an SPM image. Also the translation symmetry of the symmetrized scanning probe tip will be that of the enforced plane symmetry. It, thus, becomes quite obvious why a calibration sample should possess as high a plane symmetry group as possible in order to be of the best possible use.

To summarize the main ideas of this section: plane symmetry enforcing of the 16 higher symmetric plane groups results simultaneously in both translation averaging and point symmetry averaging for each individual pixel in an experimental SPM image. The plane symmetry enforcing procedure is the more effective the higher the plane symmetry of the imaged 2D-periodic arrays is and the more unit cells are available for the averaging. All kinds of (non-periodic) defects in the periodic motifs and noise can then be readily averaged out. However, if the plane symmetry of the sample surface features is not known in advance, the actual deviations of the SPM images from more or less truthful representations of the sample surfaces including the effects of noise have to be reasonably small for the procedure to be justifiable and meaningful.

### 2.4 Unavoidable convolution of the sample surface with an "effective" scanning probe tip and required prior knowledge to make CIP useful

The convolution of an effective scanning probe tip function, $f_{tip}(x,y)$, with the sample surface feature function, $f_{sample}(x,y)$, is unavoidable in the process of recording a raw SPM image with intensity, $i_{raw}(x,y)$. Mathematically, this convolution reads $i_{raw}(x,y) = f_{sample}(x,y) \otimes f_{tip}(x,y)$ (3a), where $x$ and $y$ stand for the real space coordinates of pixels in the 2D scanning plane, all functions are considered to be time independent, and all noise is ignored. We consider the whole SPM and its influence on the imaging process during one experiment as contributing to the effective scanning probe tip, although we do not state this explicitly throughout the remainder of this paper. One may with relation (3a) interpret $i_{raw}(x,y)$ as either being due to the tip probing the sample or the sample probing the tip. This is in compliance with the reciprocity principle of SPM imaging [3].

Provided that there is a sufficiently high number of pixels in $i_{raw}(x,y)$ so that this function can be approximated as a set of infinitely close spaced image points, we have in Fourier space the relation $I_{raw}(H,K) = F_{sample}(H,K) \bullet F_{tip}(H,K)$ (3b). When the time domain noise is negligible and decoupled form the geometrical distortions as obtainable in the limit of slow scan speeds, $f_{tip}(x,y)$ and $F_{tip}(H,K)$ represent a set of prevalent SPM operating conditions and the geometry (and possibly bluntness) of the scanning probe tip itself. These instrument and experiment specific functions can then be considered as being nearly translation invariant, i.e. approximately the same for each pixel in an image. With negligible and decoupled time domain noise (and ignoring possible point symmetries of a blunt scanning probe tips that are greater than *1*), the SPM imaging process is linear, i.e. the image of the sum of all surface features is equal to the sum of the corresponding images of the individual surface features.

This section allows for conclusions on both the utility of CIP for scanning probe microscopy and its limitations. Without some prior knowledge of either the plane symmetry of the sample surface or the shape, spatial extent, and point symmetry of the scanning probe tip as well as on the overall geometric distortions in the SPM imaging process, the convolutions of relations (3a) and (3b) can not be resolved. On the other hand, the convolutions of relations (3a) and (3b) offer the opportunity to estimate and utilize the prevalent point spread function of a SPM, as will be demonstrated experimentally elsewhere. (Note that the overall geometric distortions in the SPM imaging process "create" the "effective" tip which represents the less then ideal imaging capabilities of a particular microscope with a physical tip.)

If one is justified to assume that the scanning probe tip is close to its ideal, i.e. an infinitely sharp needle, one may enforce the most probably plane symmetry on a sample confidently and shall obtain a "more truthful" representation of the array. We will demonstrate this on suitable examples elsewhere. The prior knowledge concerns in this case both the scanning probe tip and symmetry information on the sample, where the latter may be incomplete. For example, the 2D point symmetry of molecules that form a 2D-periodic array may be broken in some unknown way, but their resulting site symmetry should in most cases be just a subgroup of the 2D point group of the isolated molecules. In special cases, the supergroup/subgroup relations may not be applicable due to the occurrence of so called "skewed symmetries" [24].

If, on the other hand, one cannot recognize the entities in the 2D-periodic array on the basis of some (possibly incomplete) prior knowledge on their shape in raw SPM images, one has to conclude that the bluntness of the scanning probe tip prevented this recognition. The purpose of applying CIP on the respective SPM images is then to symmetrize the blunt tip in order to obtain a "more truthful" representation of the array. We believe that this is the case in the experimental example of this book chapter, a STM image from a regular array of fluorinated cobalt phthalocyanine molecules on graphite, Fig. 1a.

### 2.5 Estimation and utilization of the prevalent point spread function of a scanning probe microscope

Prior knowledge of either $F_{sample}(H,K)$ or $F_{tip}(H,K)$ allows one to resolve relation (3b) for either of these two Fourier transforms by a simple division in reciprocal space, e.g. $F_{effective\ tip}(H,K) = I_{raw\ of\ known}(H,K) / F_{known\ sample}(H,K)$ (4), where we changed the subscripts in order to be more specific. With the subscript *effective tip* we mean the specific scanning probe tip that produced the raw image of a specific known sample with a direct space surface sample function $f_{known\ sample}(x,y)$. In the remainder of this section, we will continue changing subscripts for clarity and specificity.

When the prior knowledge concerns the plane symmetry of a 2D-periodic and highly symmetric calibration sample, the symmetry enforcing procedure results in a SPM image that represents the symmetry of the known sample surface features nearly truthfully, i.e. $I_{symmetry\ corrected}(H,K) \approx F_{sample\ with\ known\ plane\ symmetry}(H,K)$ (5a), with $F_{symmetry\ enforced\ tip}(H,K) \to unity$ (5b) and $f_{symmetry\ enforced\ tip}(x,y) \to \delta(x,y)$ (5c), where $\delta(x,y)$ represents a 2D Dirac delta function. Relations (5a) and (5b) are special "calibration sample versions" of relation (3b). Relation (5c) is simply the Fourier transform of (5b).

Also, relations (5b) and (5c) can be considered to referring to an ideal SPM tip that can experimentally only be approached but not reached. Relation (5a) is an idealization as well since it assumes (quite) perfectly obeyed plane symmetry for the calibration sample with many repeats of the periodic motif, negligible noise and only a minute number of defects. This sample should be (almost) free of defects that destroy the strict mathematical conditions that are set by the plane symmetry. Such defects may include thermal vibrations that are anisotropic and physisorbed surface contaminations that are inhomogeneous.

After the enforcing of the plane symmetry of the known calibration sample, one can rewrite relation (4) with (5a and 5b) as $F_{effective\ tip}(H,K) \approx I_{raw\ of\ known}(H,K) / I_{symmetry\ corrected}(H,K) = F_{PSF}(H,K)$ (6a) or $\{F_{effective\ tip}(H,K)\}^{-1} \approx I_{symmetry\ corrected}(H,K) / I_{raw\ of\ known}(H,K) = F_{invPSF}(H,K)$ (6b) and obtains a means to estimate the Fourier transform of the point spread function, $F_{PSF}(H,K)$, experimentally for a certain scanning probe tip and a certain set of experimental conditions of the SPM. The estimate for the Fourier transform of the inverse point spread function, $F_{invPSF}(H,K)$, can be used directly for correcting images of unknown samples that possess the same real space 2D periodicity as the calibration sample. These images need to be recorded with the same scanning probe tip and under the same experimental conditions either prior to or after the recording of the SPM image of the unknown. The respective relation is $I_{corrected\ of\ unknown}(H,K) \approx I_{raw\ of\ unknown}(H,K) \cdot F_{invPSF}(H,K)$ (7a), which Fourier back-transforms into an approximation of the correctly deconvoluted image in direct space $i_{corrected\ of\ unknown}(x,y) \approx i_{raw\ of\ unknown}(x,y) \otimes f_{invPSF}(x,y)$ (7b). Producing a calibration sample with the same 2D periodicity but much higher plane symmetry than the sample to be investigated may be especially useful for AFM imaging of features that are much larger than atoms or molecules.

Also, the inverse of the Fourier transform of the point spread function as determined by relation (6b) can be transformed back into direct space. There it can be convoluted with a raw real space SPM image that has been recorded with the same scanning probe tip and under the same experimental conditions. This leads as well to an approximation of the corrected SPM image of an unknown sample where geometric distortions that are due to the imaging process are largely removed.

For the whole procedure to work satisfactory, it is important that the point spread function (and its inverse function) samples the aberrations of the SPM imaging process well. This can be achieved by the usage of calibration samples that possesses many fine surface details so that the Fourier transform of its surface feature function possesses many components.

## 3. Correction of SPM images for the effects of a blunt tip

### 3.1 Correction of an experimental STM image from a 2D-periodic array of fluorinated cobalt phthalocyanine molecules ($F_{16}CoPc$) on graphite

The STM image of Fig. 1a has been recorded from a regular 2D-periodic array of $F_{16}CoPc$ molecules on highly (0001) oriented pyrolytic graphite (that is technically know as HOPG) at 20 K, at the Technical University of Chemnitz under ultra high vacuum conditions and with a tungsten tip (that was not cooled). When the periodic motif in Fig. 1a is compared to a sketch of an isolated $F_{16}CoPc$ molecule, Fig. 2a, no resemblance to the latter is found for any particular orientation of the molecule with respect to the basis vectors of the lattice. The (isolated) molecule itself possesses the 2D point symmetry *4mm*. The highest symmetric plane group that is possible for such a molecule when it remains undistorted as part of a 2D-periodic array is *p4mm*, where the positions (0,0) and (½,½) possess this site symmetry and the 2D lattice is a square.

One of two possible *p4mm* arrangement of $F_{16}CoPc$ molecules where just the (0,0) site is occupied is shown in Fig. 2b. Figure 2c, on the other hand, shows a *p4mm* arrangement of two $F_{16}CoPc$ molecules with both *4mm* sites occupied.

The lattice remains primitive, i.e. there is no centring as indicated by the leading *p* in the plane group symmetry symbol, but two molecules need to "pair up" to form a new periodic motif that fills the whole plane by translations, Fig. 2c. For all three theoretically possible 2D-periodic arrangements of undistorted $F_{16}CoPc$ molecules, no resemblance to the periodic motif in the raw STM data, Fig. 1a, can be discerned. While the unit cell of these three possible 2D-periodic arrangements is a square, the unit cell of the raw STM image, Fig. 1a, is oblique.

We strongly suspect, therefore, that the raw image of Fig. 1a has been recorded with a blunt tip. This blunt tip may consist either of protruding orbitals from many individual W atoms or from whole $F_{16}CoPc$ molecules that are attached to the tip, or perhaps a combination of both of these possibilities. The periodic motif in this figure is, thus, probably more representative of a blunt scanning probe tip [1, 25] than of genuine sample surface features, which are assumed to be (negligibly disturbed) individual $F_{16}CoPc$ molecules that are arranged in a 2D spatially periodic manner.

It is also well known from simulations that multiple mini-tips tend to obscure vacancies at the periodic motif level in SPM images of 2D-periodic surface features [25]. No such vacancies are visible in the raw STM image of Fig. 1a, suggesting that this image was indeed recorded with a blunt STM tip.

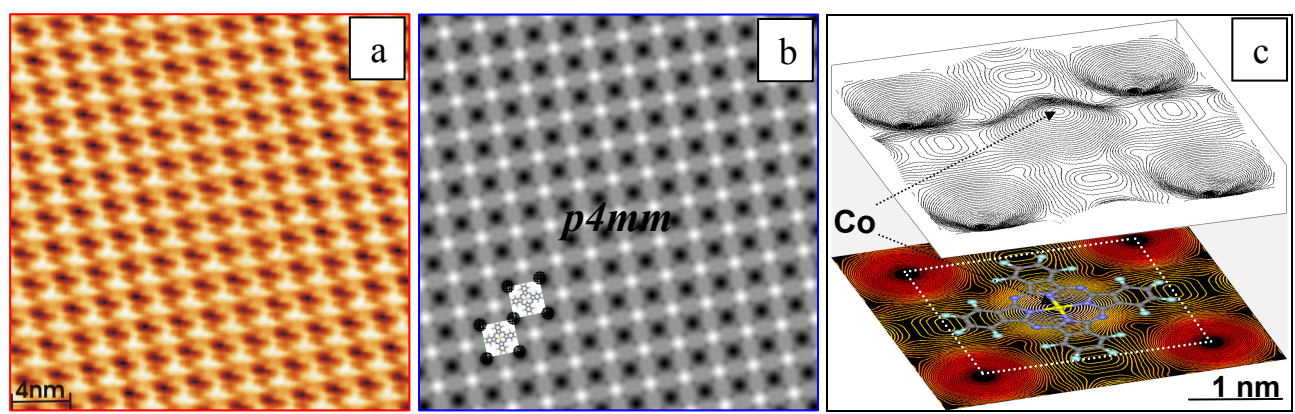

**Fig. 1** STM images of a regular 2D-periodic array of $F_{16}CoPc$ molecules on HOPG. **a)** *Raw data*, 1771 by 1771 pixels, constant tunnelling current mode, tip bias 1 V (with respect to the more negative sample), 0.1 nA tunnelling current. **b)** *p4mm symmetry enforced version* of the raw data with sketches of two molecules as inset. **c)** Contour plots of 1.5 unit cells of the *p4mm* enforced data with one molecule sketched in and one unit cell boxed by dotted lines. The contour plots convert the intensity distribution of (b) into 64 levels. While the four very deep depressions in c) that surround each molecule are darkest in (b), the local density of electronic states is highest above the central Co atom, which is brightest in (b). The *p4mm* symmetry enforcing procedure can be thought of as aligning the periodic motifs of all independent STM images from the multiple mini-tips on top of each other, thus, enhancing the signal to noise ratio significantly. (Modified after refs. [10, 11]).

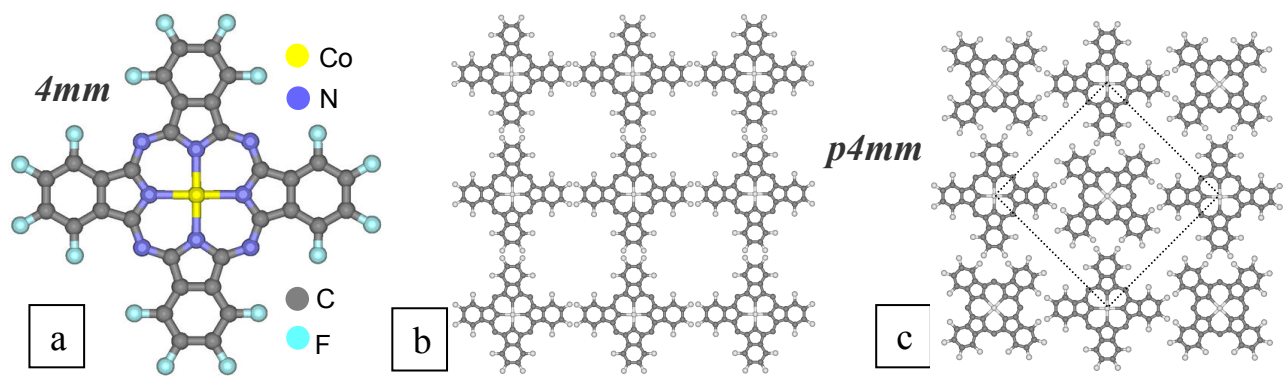

**Fig. 2 a)** Model of an isolated molecule of $F_{16}CoPc$ and two possible 2D-periodic arrangements of this molecule in plane group *p4mm* under the assumption that the molecule remains geometrically undistorted **b)** and **c)**. While b) shows a 2D-periodic arrangement of this molecule with a "single molecule" motif, c) required a "paired-up molecules" motif. The 2D point symmetry group of all molecules in this figure is *4mm*. The 2D lattice is in both 2D-periodic arrangements of the motifs a square. One unit cell is boxed by dotted lines in c. While nine molecules represent an area of 9 (primitive) unit cells in b), the same number of molecules occupies an area of only 4.5 (primitive) unit cells in c). (Modified after refs. [10, 11]).

Using the program CRISP [8], a central circular section with a diameter of 1024 pixels of Fig. 1a was processed crystallographically and plane group *p4mm* enforced on the raw data. The result is shown as Fig. 1b at about the same magnification as the raw data, Fig. 1a. Note that ***neither*** information on the nature of the atoms ***nor*** on their relative position within the molecules (nor on the nature of the STM imaging process) was employed during the CIP procedures. All we introduced as prior knowledge was just the abstract 2D point symmetry of isolated $F_{16}CoPc$ molecules and the entirely reasonable assumption that this symmetry is not severely broken when the molecules become part of the 2D-periodic array on weakly interacting HOPG. The clearly visible result of CIP on the raw data, Fig. 1a, i.e. that one can now discern the individual $F_{16}CoPc$ molecules in Fig 1b is, therefore, all the more remarkable! In effect, the symmetrized image of Fig. 1b shows the idealized electronic structure of the molecular $F_{16}CoPc$ array (in bright features of varying intensity) on HOPG (that forms the dark background and shows up as 4 very dark "holes" around each molecule, Fig. 1c). This figure also shows that the single-molecule motif arrangement of Fig. 2b is realized for geometrically undisturbed $F_{16}CoPc$ molecules that seem to form a regular 2D array with a square unit cell on HOPG. Applying the reciprocity principle of SPM imaging, i.e. *"an image of microscopic scale may be interpreted either as by probing the sample state with a tip state or by probing the tip state with a sample state"* [3], the emergence of clear representations of $F_{16}CoPc$ molecules in Fig. 1b can be interpreted as being due to the symmetrizing of a blunt STM tip.

It must be clearly stated here that the plane symmetry of the true natural arrangement of the $F_{16}CoPc$ molecules on HOPG is **not** proven to be *p4mm* by the application of the CIP procedures. This is something that these procedures are currently not capable of delivering[7]. To be fair, this is a question that we did **not** promise to answer in this book chapter. When one is concerned with the symmetrization of a blunt STM tip and an associated improvement of the signal to noise ratio, obtaining a good approximation of the image of the average molecule is all that matters (and this has indeed been achieved, see Fig. 1b).

What we can say with certainty (on the basis of space symmetries that are possible in 2D) is that when the $F_{16}CoPc$ molecules are undistorted in the 2D array and when their site symmetry (within the array) coincides with their point symmetry, plane group *p4mm* is required for their mutual arrangement. If, on the other hand, the point symmetry of the molecules were distorted to a subgroup of *p4mm* (due to them being part of the array) and if the molecules were located at positions of site symmetry that match their reduced point symmetry, one could interpret the net result of *p4mm* plane symmetry enforcing as the "recovery" of a geometrically undistorted image of the average $F_{16}CoPc$ molecule.

Because STM images are approximately maps of the local density of electronic states (LDOS) of the sample surface at the Fermi level convoluted with the electronic structure of the scanning probe tip [3], the "pure geometry" of an atomic arrangement is not necessarily revealed in a STM image, see Fig. 1c, since the LDOS may be spread over a whole molecule [4]. Depending on the sign and amount of the sample bias, the STM images of physisorbed $F_{16}CoPc$ molecules on HOPG may reveal approximations of the lowest unoccupied molecular energy orbital or the highest occupied molecular energy orbital, both modified by electronic coupling effects to the graphite [26].

By far the highest tunnelling currents were observed by other authors directly above the central Co atoms in more or less quadratic arrays of CoPc molecules on the (111) surface of gold[8], see refs. [26, 27]. Because we see similar effects in Figs. 1b and 1c (for a homologous molecule where F atoms substitute for H atoms), we take our independent observation of 2D periodic, highly localized maxima of the tunnelling signal at the positions of the Co atoms as confirmation of the validity of our *p4mm* symmetry enforcement on the raw image of Fig. 1a. (As demonstrated in refs. [10, 11], the enforcing of subgroups of *p4mm* on the raw STM image of Fig. 1a does not result in images with a regular array of pronounced intensity maxima that can be interpreted straightforwardly as the location of Co atoms.)

Mizes and co-workers [28] proposed that two acting STM mini-tips will tend to produce just a linear superposition of two independent STM images. A corresponding effect of the "apparent doubling" of widely (i.e. about 0.8 nm) separated (ad)atoms in the 7 by 7 reconstruction of Si (111) when recorded with two STM mini-tips has been observed experimentally [29]. Jack Straton of Portland State University re-examined the work of Mizes and co-workers on the basis of first principle calculations for the STM image contrast employing linear combinations of atomic orbitals[9].

On the basis of our analyses[9], we are quite certain that the effects of two acting mini-tips can be straightforwardly generalized for multiple mini-tips that scan 2D-periodic surface features. Neglecting interference effects[9], a blunt STM tip that consists of multiple mini-tips will, thus, record many independent images of the sample surface features. The linear superposition of all of these images will result in an image of the blunt STM tip as probed by the array of surface features. The following section will illustrate this and the symmetrizing effects of CIP on the basis of simple simulations of regular 2D-periodic arrays of bright crosses on a dark background.

---

[7] When multiple mini-tip effects are involved and prior knowledge on the spatial distribution of the mini-tips (i.e. the point symmetry of the blunt tip in other words) is not available, providing such a proof may be extremely difficult even for refined CIP procedures. Also, it might be possible that the unknown peculiarities (i.e. approximate point symmetry higher than *1* and specific atomic structure) of the mini-tips bias the results of CIP towards a particular mutual arrangement of the surface features that is simply wrong. One may encounter such erroneous analyses particularly often when a blunt tip possesses a point symmetry that is not also a site symmetry of the 2D-periodic array of surface features and prior knowledge about this fact was lacking. It may, therefore, be possible that the "true natural" arrangement of geometrically distorted $F_{16}CoPc$ molecules on HOPG possesses only plane symmetry *p4*, but that "approximate" *m* or *2mm* point symmetries of the blunt tip biased this arrangement towards the CIP results of Fig. 1b.

[8] From the viewpoint of the plane symmetry of the substrate and the molecular point symmetry alone, the deposit/substrate combination of refs. [26, 27] is equivalent to $F_{16}CoPc$ molecules on a (0001) oriented graphite surface. Existing differences in the size of the unit cells, possible interactions between the molecules themselves and with the substrate as well as possible surface reconstructions of the gold may all be ignored.

[9] Our main conclusions from that study were that two acting STM mini-tips will indeed produce terms that are linear superpositions of two independent STM images. In Fourier space, these two independent images will differ only in a phase angle term. The CIP procedures simply remove this phase angle term and the results are the superposition of the two independent STM images and a corresponding enhancement of the signal to noise ratio of the imaging process. The same is true for imaging with more than two acting STM mini-tips, which results in multiple independent images. There are also interference effects between two (or more) mini-tips that represent small systematic deviations from the linear superposition of two (or more) independent STM images. Supporting simulations, to be published elsewhere, show, however, convincingly that CIP does work satisfactory for the vast majority of "blunt STM tip images" despite the presence of interference effects. When interference effects and more than two mini-tips are present, their periodicities are unlikely to be the same as those of the surface features. CIP will either filter the interference effects out (just as it does with non-periodic noise) or symmetrize them (i.e. average them in a symmetric manner). We are, therefore, convinced that both multiple mini-tips and associated interference effects do not cause problems that invalidate the utility of CIP for STM.

### 3.2 Supporting simulations that ignore interference effects between multiple SPM tips for simplicity

A hypothetical sample surface that possesses cross-shaped features with point symmetry *4mm* which are arranged in a regular array with a square unit cell possesses plane symmetry *p4mm*. For simplicity, we consider only a "one cross" equals "one periodic motif" arrangement in plane group *p4mm*, see, e.g., Fig. 2b. When this sample is accessed by a SPM with a single scanning probe tip, one obtains a SPM image with a plane symmetry that is very close to *p4mm*. (No corresponding image is shown in this book chapter. We concern ourselves with such a hypothetical image here only as being a single component of a raw SPM image that shows the effects of multiple scanning probe tips).

When this hypothetical sample is scanned by an asymmetric[10] blunt SPM (that lacks all point symmetries higher than *1*) and is composed of two, three, or more scanning probe mini-tips, one obtains a SPM image which consists of multiple image components. Since the SPM imaging process is (under the above mentioned conditions) linear, multiple images of the same periodic sample surface will be superimposed and form a combined image where the 2D-periodic image motif is a two, three, or more-fold superposition of the sample surface features. Besides the identity element (*1*), the plane symmetry group of the resulting SPM image will then only possess translation symmetry[11].

Figures 3 and 4 illustrate the results of CIP (using the program CRISP [8]) on two simulated 2D-periodic "raw images" that were created in order to mimic multiple scanning probe tip effects. Each of these images can be interpreted as being the sum of three independent images of a single quadratic array of crosses. Figures 3a and 4a show the two raw images and differ in the nature and mutual arrangement of the individual crosses. Note that each of the two raw images possesses only one kind of cross (i.e. either an x or a +) since this is a condition for being interpretable as a composite image. These two kinds of crosses differ in their geometry, but both possess point symmetry *4mm*.

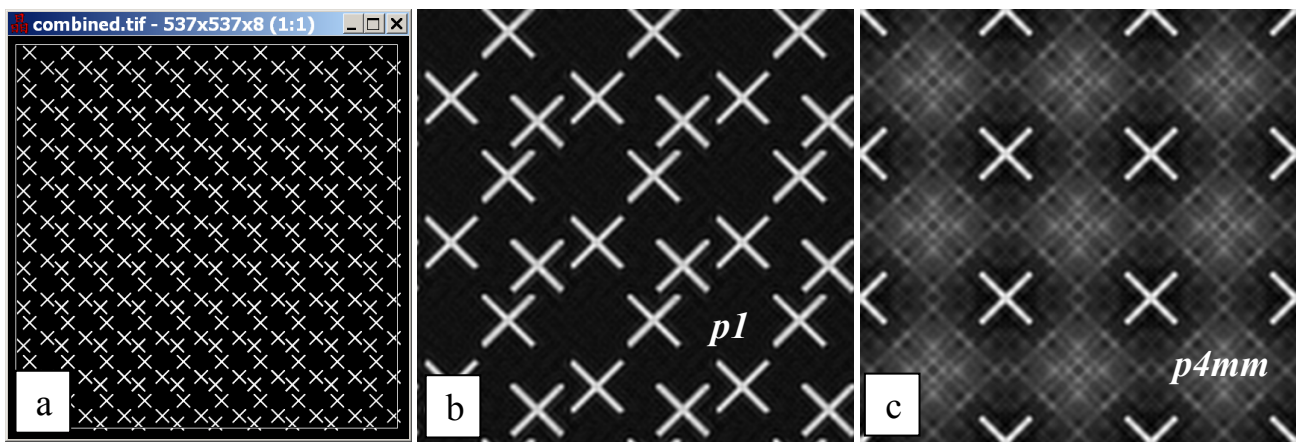

**Fig. 3** CIP on a simulated image to illustrate how multiple SPM tip effects are dealt with as a matter of principle. **a)** is the raw image that was created by linearly superimposing the same quadratic array of crosses three times with differing lateral shifts. While **b)** shows a 9 unit cell section of the translation averaged version of the raw image, **c)** shows a 9 unit cell section of the *p4mm* enforced version of this image. Note that a 512 by 512 pixel quadratic section of the raw image has been selected for the application of the CIP procedures. (Using a 512 pixel diameter circular section instead has no significant influence on the results of the CIP procedures.)

The plane symmetry of the two raw images, Figs. 3a and 4a, is, as expected, only *p1* because we avoided any special mutual arrangement of the three scanning probe tips. The unit cells of both raw images contain three crosses as the periodic motifs and the asymmetry, i.e. point group *1*, of the motifs remains so after the *p1* enforcement. As expected, enforcing plane symmetry *p1* (i.e. only translation symmetry) on the raw images does not change the "appearance" of these images significantly, Figs. 3b and 4b. With appearance we mean here mainly the distribution and nature of the "three cross motifs". Since the three crosses remain essentially unchanged, one can also say that the scanning probe tip has not been symmetrized by our enforcing of translation symmetry, which existed of course before in the raw images.

---

[10] As long as there is no "special" mutual arrangement of the multiple scanning probe mini-tips that result in a point symmetry higher than *1*, no additional site symmetries can be created and existing site symmetries cannot be modified by imaging with such a blunt scanning probe tip.

[11] The unit cell size of the combined image will be the same as that of all component images (since the linear superposition does not change the periodicity). The unit cell shape can be considered to change in correspondence with the lowering of the point symmetry of the superimposed motif. Under the above stated "asymmetry condition" for the blunt scanning probe tip, there is an infinite number of oblique parallelograms that can all be considered as representing the unit cell of the combined SPM image. Since squares may be considered as being very special cases of oblique parallelograms, the (underlying) square unit cells of the three single cross arrays are also possible unit cells of the combined image.

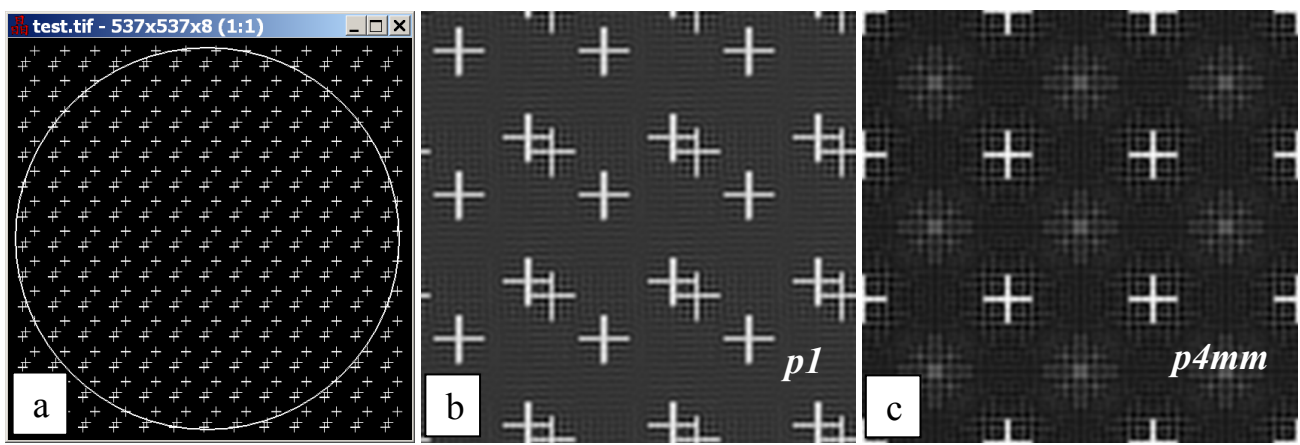

**Fig. 4** CIP on another simulated image to illustrate how multiple SPM tip effects are dealt with. **a)** is the raw image that was created by linearly superimposing the same quadratic array of crosses three times with differing lateral shifts. While **b)** shows a 9 unit cell section of the translation averaged version of the raw image, **c)** shows a 9 unit cell section of the *p4mm* enforced version of this image. Note that a 512 pixel diameter circular section of the raw image has been selected for the application of the CIP procedures. (Using a 512 by 512 pixel quadratic section instead has no significant influence on the results of the CIP procedures.) Note also that Fig. 4a presents a challenge to any Fourier processing algorithm because many Fourier coefficients are needed in order to reproduce contrast at sharp edges that changes with very high spatial frequencies.

While CIP processing artefacts become visible in Fig. 4b, they are negligible in Fig. 3b. The artefacts in Fig. 4b are partly due to the nature of the three cross motif, Fourier processing, and the fact that a finite section of the periodic array needed to be selected for the CIP procedure. We did choose a black background on purpose for both raw images in order to make these kinds of artefacts visually more outstanding.

Enforcing *p4mm* for the position about which this symmetry was the least broken leads to the symmetrized versions of the raw images. These versions/images are shown as Figs. 3c and 4c with the same magnification as the *p1* enforced images. As expected, these two images do possess a significantly different appearance with respect to the raw and *p1* enforced images. Instead of the "three cross motifs", there are now only "one cross motifs" and some CIP processing artefacts. These artefacts arise again partly from the need to select a finite patch of a theoretically infinitely large image of a 2D-periodic array. As one would expect, the shape of the unit cell is a square in Figs. 3c and 4c.

One may, thus, consider the symmetry enforcing step of the CIP procedure as being equivalent to shifting intensity in the images to the right positions where a single SPM tip would have detected it, i.e. to the crystallographic origin and its translation equivalent positions, while making sure that the intensity distribution around each of these positions possesses the required site symmetry. For plane group *p4mm*, the site symmetry at the origin is *4mm*. Since this is also the point symmetry of each of the simulated crosses, one may say loosely that per unit cell one of the crosses remained at the same position while the other two crosses were brought into coincidence with it. Employing the reciprocity principle of SPM [3] and Pierre Curie's symmetry principle [5, 23], this is equivalent to shifting the three independent scanning probe mini-tips into coincidence and symmetrizing the resulting single tip to point group *4mm*. (As one would expect, the enforcing of subgroups of *p4mm* on Figs. 3a and 4a does not result in the superpositions of the three crosses. This is easily understood as none of these subgroups contains positions with site symmetry *4mm*.)

**Acknowledgements** Support by Portland State University "Venture Fund", "Research Stimulus", "Faculty Enhancement", and "Internationalization" awards is acknowledged. Prof. Wolfgang Neumann of Humboldt University Berlin is thanked for both the critical proof reading of the manuscript and his hospitality as sabbatical host during the time of the writing of this book chapter. Dr. Ines Häusler of Humboldt University Berlin is thanked for creating Figures 1c and 2. Fruitful discussions with Prof. Michael Hietschold of Chemnitz Technical University and his students are also acknowledged. The collaborative contributions of Prof. Jack C. Straton to this project at Portland State University are also very much appreciated.